# Taxonomy of defects in semi-dry transferred CVD graphene

N. Reckinger[1,*] and B. Hackens[1]

[1]Université catholique de Louvain (UCLouvain), Institute of Condensed Matter and Nanosciences (IMCN), Nanoscopic Physics (NAPS), Chemin du Cyclotron 2, 1348 Louvain-la-Neuve, Belgium.

* Corresponding author: E-mail address: nicolas.reckinger@uclouvain.be, nicolas.reckinger@gmail.com.

## Abstract

Post-transfer in-depth morphological characterization of graphene grown by chemical vapor deposition (CVD) is of great importance to evaluate the quality and to understand the origin of defects of the transferred sheets. Herein, a semi-dry transfer technique is used to peel off millimeter-sized CVD graphene flakes from polycrystalline copper foils and transfer them onto SiO2/Si substrates. We take advantage of the unique feature of this semi-dry process: it preserves the copper substrate, enabling location-specific morphological comparisons between graphene and copper at various stages of the transfer. Thanks to a combination of morphological characterization techniques, this leads to trace and elucidate the origin of various post-transfer graphene defects (cracks, wrinkles, holes, tears). Specifically, thermally induced wrinkles are shown to evolve into nanoscale cracks, while copper surface steps lead to folds. Furthermore, we find that the macroscale topography of the copper foil also plays a critical role in defect formation. This work provides guidelines on how to correctly interpret the post-transfer morphology of graphene films on relevant substrates and how to properly assess their quality. This contributes to the optimization of both the graphene CVD growth and transfer processes for future applications.

## 1) Introduction

The growth of graphene by chemical vapor deposition (CVD) on metallic catalysts is by far the most popular technique for the production of high-quality, large-scale graphene sheets. Since its inception in 2009,[1] the CVD growth of graphene on copper substrates has elicited a significant number of publications and known tremendous progress.[2,3] However, under mainstream growth conditions with commercial copper foils, it remains highly difficult to obtain an ideal, *i.e.* defect-free and flat, graphene film. These intrinsic (*i.e.* originating from the CVD growth) imperfections[4] include domain boundaries, adlayers,[5,6,7,8,9] wrinkles,[10,11,12,13] contaminations,[14,15,16] cracks,[17,18,19] *etc.* In addition, since graphene on copper alone has very limited interest, it must be transferred to other substrates for most applications. Like CVD growth on copper, the transfer process itself introduces its share of defects, the main ones being wrinkles/folds/pleats and cracks/tears.[20,21] Both the intrinsic and extrinsic (*i.e.* arising from the transfer) defects have a profound impact on the final quality of graphene. In the specific case of wrinkles, and more particularly cracks, it can be challenging to attribute their origin to either the growth or to the transfer processes.

Thermally induced wrinkles are well documented in the literature and easily recognizable with optical microscopy (OM) and scanning electron microscopy (SEM). They are induced during the cooling step following the graphene CVD growth, owing to compressive strain arising from the difference in thermal expansion coefficients between graphene and its growth substrate.[22,23,24] One research group has shown that some wrinkles can be also induced by the morphology of the copper foil (copper steps) and that it is possible to distinguish those from the thermally induced ones.[11,12] Wrinkles are usually classified into three types,[13] according to their geometrical and morphological characteristics: (1) ripples (extremely narrow (~1 nm) and with a low aspect ratio around), (2) standing collapsed wrinkle (broad range of heights and very narrow), (3) folded wrinkles, being wide (broad width distribution, typically between 20 and 150 nm or more) and flat (typical height of ~1 nm). Based on energetic considerations, the wrinkles of type (2) and (3) are in fact an evolution of the simple ripple dictated by the height: a too high ripple results in a standing collapsed wrinkle while a folded wrinkle arise from the folding of a too high standing collapsed wrinkle. On the other hand, cracks observed in graphene after CVD growth are barely reported on and their formation mechanism is not well understood.[17,18,19] As opposed to thermally induced wrinkles caused by compressive stresses exerted during the cooling, intrinsic cracks in graphene on copper must in all likeliness result from a tensile stress applied in a direction perpendicular to the crack.[18,25]

Dry-transfer methods for CVD graphene grown on copper have been developed for the purpose of avoiding as much as possible the contamination of graphene that is inevitable with the widely used wet transfer. They can be subdivided into the *all-dry*[26,27,28,29,30] and *semi-dry*[30,31,32,33] categories. The all-dry transfer means, by definition, that graphene is not exposed to any liquid during the process, either to separate it from copper (condition 1) or to remove the polymer support mandatory for achieving mechanical graphene lamination onto the target substrate (condition 2). For instance, it was used to transfer large-scale, complete graphene films with a polymeric scaffold[26,27] or, drawing inspiration from methods developed for exfoliated 2D materials,[34,35,36] to fully protect graphene by encapsulation between two hexagonal boron nitride (hBN) flakes.[28,29,30] In contrast, the semi-dry flavor releases one the two conditions of the all-dry transfer. Hu et al.[31] peeled off wafer-scale graphene from copper with a stack of polymers that were dissolved in an adequate solvent (breaking condition 2). On the other hand, hBN encapsulation of graphene was achieved by picking up graphene transferred beforehand onto $SiO_2$/Si by traditional wet etching (breaking condition 1).[32,33] Moreover, a successful mechanical, dry delamination of CVD graphene away from the copper surface requires to weaken their mutual interaction. Decoupling graphene from copper can be accomplished by oxidizing copper at the interface in different ways: naturally in ambient air,[28,37] in saturated water vapor,[29,30,38] by immersion in room-temperature water,[26] hot water,[37,39] or a mixture of water and ethanol.[27,31] Excellent results in terms of oxidation efficiency and speed are obtained with the last technique.

Here, we employ a semi-dry transfer technique to delaminate large-area graphene flakes from polycrystalline copper foils and transfer them on $SiO_2$/Si substrates. The observation of the resulting transferred graphene reveals different kinds of nanoscale linear defects in the form of cracks and wrinkles, as well as larger scale holes and tears. The main objective of this paper is to give an honest account of these imperfections and elucidate their origin (growth and/or transfer), first step toward addressing them. To do so, samples are scrutinized at different stages of the transfer process at the same location using a combination of morphological characterization techniques. Depending on their type and density, they will limit accordingly the area of application of transferred graphene samples. It is therefore very important to be able to properly qualify these defects, as well as to define different quality grades of CVD graphene.[40] This is very useful for helping end users to determine what grade of graphene fits best their target application. For example, a few narrow cracks in graphene would not be disqualifying for some applications but certainly for solid-state electronics. This work will likely help researchers striving to improve the quality of their transferred CVD graphene.

## 2) Methods

**Fabrication**

*Graphene CVD growth* Graphene is grown on copper foils electropolished with a home-made setup, following a CVD process described in details in a previous paper.[16] In short, once the sample is inserted, the fused silica reactor of the hot-wall CVD furnace is pumped down to primary vacuum (~$10^{-2}$ mbar) and then refilled with Ar up to atmospheric pressure to remove as much ambient air as possible. Next, the temperature of the furnace is increased to 1050 °C and Ar alone (with ~1 ppm of residual $O_2$) is flown to decrease the nucleation density and increase the grain size by depleting carbon present in as-received copper foils. After 90 min, the sample is reduced for 20 min by adding Ar/$H_2$ to Ar. The growth itself is then performed by adding Ar/$CH_4$ for 60 min to grow isolated monolayer millimeter-scale graphene flakes. Finally, the furnace cools down naturally to room temperature after shutting down the power supply, with the same mixture of gases flowing in the reactor.

*Graphene semi-dry transfer* The graphene/copper substrates are immersed in a 1:1 ethanol/deionized (DI) water mixture for at least 10 h to decouple copper and graphene by oxidizing the copper in direct contact with graphene.[31] On the other hand, the stamp used for the transfer comprises a piece of polydimethylsiloxane (PDMS) (2 mm disk obtained with a punch in a Gel-Pak PF film with a thickness of 6.5 mil *i.e.* ~165 µm) covered by a 1-2 mm wide stripe of commercial poly(vinyl alcohol) (PVA) film, supported on a white paper sheet (with a thickness of ~40 µm).[26] After detaching PVA from its support, it is positioned on the PDMS disk at the edge a microscopy glass slide and fixed with Kapton tape. Before transfer, the graphene/copper pieces are stuck on a silicon piece, also with Kapton tape. The stamp-based semi-dry transfer is performed with a home-made transfer setup and proceeds as follows (see the protocol in Fig. 1). In order to facilitate the identification of isolated flakes on copper, the sample is heated up on a hot plate at 150 °C for a few minutes beforehand to induce mild oxidation of the bare copper surface and induce an optical contrast with graphene, without provoking the creation of defects.[41,42] The stamp is first pressed on the zone of interest on the graphene/copper sample. Then, the temperature of the chuck is increased to 105 °C to soften the PVA (with a glass transition temperature of ~80 °C)[43] and promote its contact to graphene, during a few minutes. After cooling down, the graphene flake is picked up. The PVA stripe is next delicately detached from the glass slide and positioned again on the PDMS disk, sticking by electrostatic interactions without Kapton tape this time. Finally, the graphene flake is transferred onto $SiO_2$(285-nm-thick)/Si substrates by the same sequence of steps: (1) pressing, (2) heating up at 105 °C, and (3) cooling down. The glass slide is then lifted up and the graphene/PVA stack remains stuck on the substrate. PVA is finally dissolved in DI water at room temperature for at least 24 h[26] or at 80 °C for 6 h.

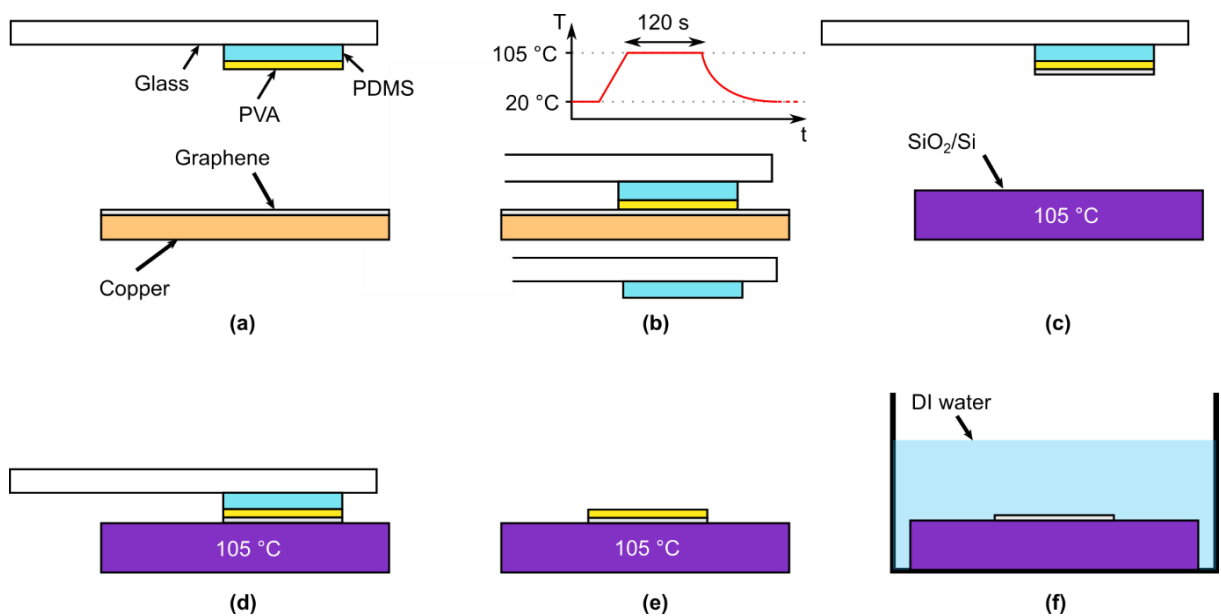

**Figure 1:** Semi-dry transfer process of CVD graphene. (a) PVA/PDMS stamp on a microscopy glass slide and CVD graphene on copper. (b) Contact between the stamp and graphene followed by heat-up. (c) and (d) Approach, contacting and heat-up of the peeled-off graphene on the SiO₂/Si substrate. (e) Removal of the glass slide and PDMS stamp. (f) Stripping of the PVA film in DI water.

**Characterization techniques**

A Zeiss Axio Imager Vario microscope was used to performed OM. The SEM images were recorded with a Zeiss Ultra 55 microscope in in-lens detector mode at an electron energy of 2 keV. Atomic force microscopy (AFM) was carried out with a Dimension Icon from Bruker in tapping mode. The images were recorded in 256 × 256 or 512 × 512 px depending on the desired accuracy and the scanning rate fixed to 0.75 (for an image size above $10 \times 10 \ \mu m^2$) or 1 Hz (for an image size below $10 \times 10 \ \mu m^2$). The data were analyzed with the Gwyddion software.[44] Raman spectroscopy (RS) is performed at room temperature with a LabRam Horiba spectrometer at a laser wavelength of 514 nm with a 2400 lines per mm grating. The laser beam is focused on the sample with a 100× objective (NA = 0.95) and the power is kept below 1 mW. The integration time is either 1-2 min or 15 s for spectrum acquisition on copper or SiO₂/Si, respectively.

## 3) Results and discussion

Initial observations reveal that the final result of the semi-dry transfer of CVD graphene on SiO₂/Si presents various types of defects (cracks, wrinkles, folds, holes, tears, *etc.*). Even if these observations have been previously reported, the present work distinguishes itself by the high degree of detail with which samples are inspected at different steps of the transfer process to diagnose the origin of these defects. Notably, contrary to wet transfer, semi-dry transfer presents a distinctive advantage: the copper substrate is not dissolved and remains thus available for post-transfer inspection. This enables correlation between the defective structures in transferred graphene and features of the surface of the copper foil after graphene pickup.

### 3.1) Nanoscale defects

The representative graphene region (~1 × 1.2 mm$^2$) that is chosen for semi-dry transfer onto SiO$_2$/Si is exhibited on copper (before wet oxidation) in Fig. S1(a) while the result of the semi-dry transfer is shown in Fig. S1(b). Even though it is in reality a cluster of two larger monocrystalline graphene flakes and two smaller ones, it is hereafter referred to as the graphene flake. In order to get more insight into the fate of graphene after transfer at the nanoscale, we have inspected in great detail a representative small area of this graphene sample at different stages of the transfer process by OM, SEM and AFM, in a manner similar to Ref.[19]: on copper after growth, on copper after wet oxidation, on copper after transfer, and on SiO$_2$/Si after transfer. Note that this sample was not annealed under HV after transfer.

Let us now observe the graphene area of interest after transfer onto SiO$_2$/Si. In Fig. 2(a), a SEM picture at moderate magnification after transfer to SiO$_2$/Si reveals different types of features displaying a bright SEM contrast. We identify three kinds of light-contrasted linear features worth of investigation in Fig. 2(a) (spotted with colored dots), based on their comparison to the corresponding SEM picture on exposed copper after transfer, displayed in Fig. 2(b): green dotted lines which can be matched to dark-contrasted features (case 1); red dotted lines which correspond to a trace displaying a lighter contrast than green dots features (case 2); blue dotted lines which coincide with apparently nothing (case 3). Figure 3(a) shows a low magnification SEM picture (it is highlighted by a red frame in Fig. S1(a)) of graphene on copper including the zone of interest of Fig. 2 (black frame). Green, red and blue frames locate the smaller regions that will be scrutinized later on in Figs. 7-9 to identify each of the three cases of linear defect after transfer onto SiO$_2$/Si.

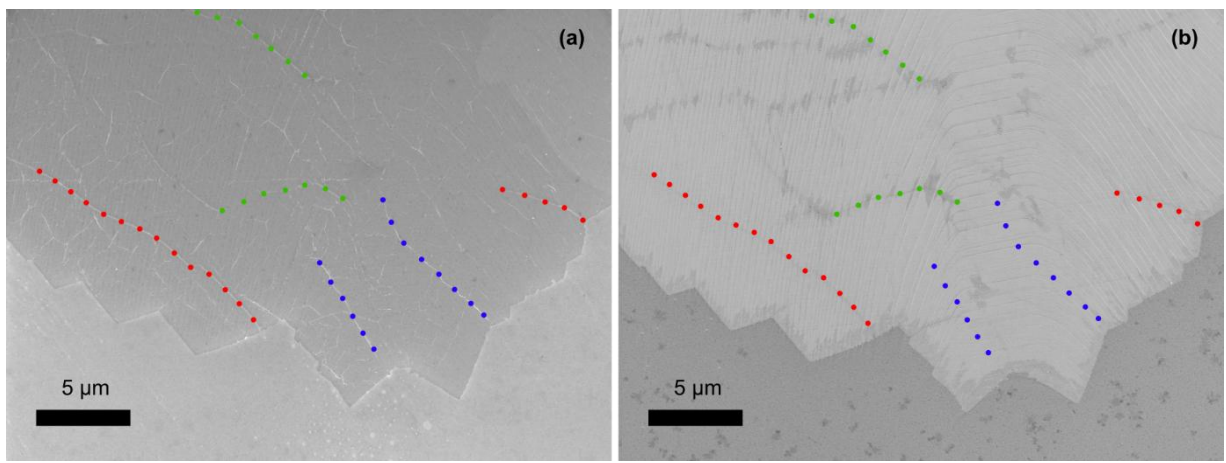

**Figure 2:** SEM pictures of the graphene zone of interest at moderate magnification (a) after semi-dry transfer onto SiO$_2$/Si and (b) on copper after stripping off graphene. The green, red and blue dotted lines indicate different types of linear defects in graphene.

Before doing that, we are now going to conduct several experiments to describe in detail and establish the nature of the numerous linear structures appearing with a dark contrast in Fig. 3(a).[19,37,48] When the same area is inspected by OM (at 100× magnification) on copper after transfer (see Fig. 3(b)), we see that the linear defects appear as reddish, what can be linked to the color of copper oxide of a given thickness.[37] The same observation can be made along the edge of the graphene flake. Anterior studies have indeed evidenced enhanced corrosion of copper in the presence of graphene at the place of defects[45,46] and also preferential copper oxidation along the graphene island edges, the oxidation progressing from the edge inward.[37] The same research paper[37] showed in addition that the composition of that copper oxide corresponds to cuprous oxide ($Cu_2O$).[47] We have performed RS on one of these red traces and found the same composition (see Fig. S2). It can thus be concluded that the dark SEM contrast observed along these linear defects is indeed related to copper oxidation. This does not come as a surprise since the sample was stored in cleanroom air (the temperature is 21.5 °C and the humidity kept below 55%) for 40 days before SEM imaging. This oxidation actually appears quite rapidly, already after 1-2 days of storage in cleanroom air (see Fig. S3(a)), and continues to evolve with time (see Figs. S3(b-d)). For copper to get oxidized, it is self-evident that it must somehow be in contact with oxidizing species. It was demonstrated previously that, in water-saturated air, it is the oxygen contained in water, and not diatomic oxygen from air, that participates in the oxidation of copper at the graphene interface[38] and the situation seems to be the same in ambient air.[37]

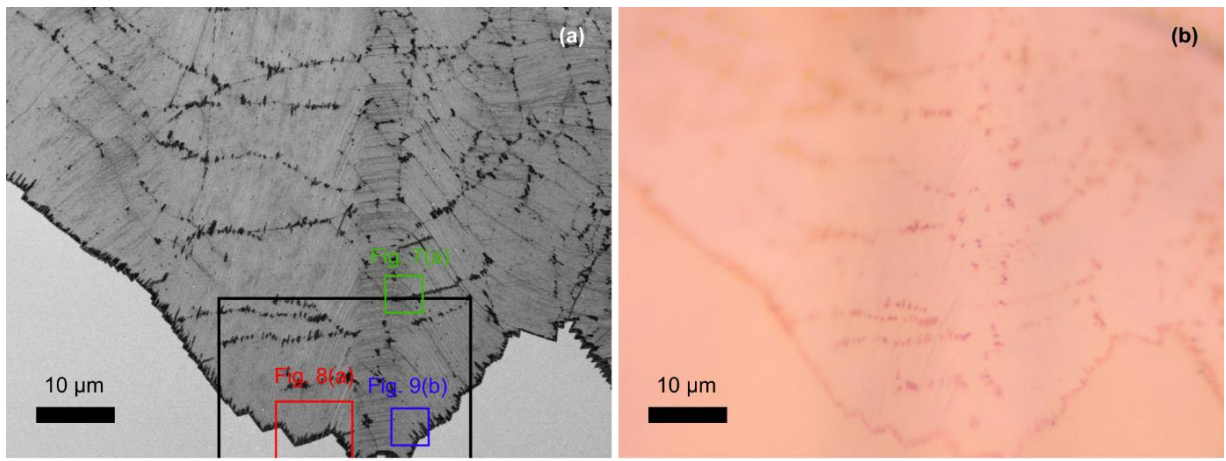

**Figure 3:** (a) Low magnification SEM picture of graphene on copper encompassing the zone of interest of Fig. 2 (black frame). The green, red and blue frames locate the regions that will be analyzed in details to identify each three types of linear defect in graphene after transfer onto $SiO_2$/Si. (b) Corresponding OM image on copper exposed after stripping off graphene.

A reasonable hypothesis is that these dark-contrasted defects may be narrow cracks in graphene, providing favorable conditions for the corrosion process to occur. If we zoom in on one of these lines after CVD growth, on copper (see Fig. 4(a)), a dark-gray line standing out from a monolayer graphene background can be distinguished, surrounded by oxidized copper regions darker still. A darker SEM contrast means the reflection of fewer secondary electrons, suggesting that these lines are composed of thicker graphene. Since contrast in SEM can sometimes be misleading (as illustrated by the contrast inversion between copper and $SiO_2$/Si in Fig. 2), a 5 × 5 µm² AFM scan was also performed on a few of these structures, after wet oxidation. In Fig. 4(b), a SEM view of a selected region is shown, together with the corresponding AFM scan (red frame) and a line profile of one of the linear structures (blue dotted line). The profile reveals that the line in question is approximately 2 nm high and 80 nm wide. This changes the scenario in favor of graphene narrow, intrinsic thermal wrinkles in place of nanoscale cracks.[37,48] If one considers the wrinkle taxonomy introduced by Ref.[13], these wrinkle dimensions imply that it should be a folded wrinkle comprising three graphene layers. However, 2 nm being thicker than three graphene layers (~1 nm), this must mean that copper is weakly oxidized just under the wrinkle. This is testified in Fig. S4 where a ~2 nm high bulge is observed at the place of a narrow wrinkle (after peeling off graphene).

A complementary experiment is performed to further confirm, if necessary, the association between oxidized copper nanocrystals and narrow thermal wrinkles by a simpler and more convenient method than AFM. The experiment consists in exposing graphene on copper to a gentle oxygen plasma to etch selectively the monolayer graphene background.[12] In this way, if narrow wrinkles are indeed present, they should be somehow preserved after the plasma since they are thicker. Figures 4(c) and (d) compare the same location before and after plasma etching by high magnification SEM imaging. Before etching, the wrinkle is barely visible while it stands out after etching. In Fig. 4(e), several "plasma-isolated" narrow wrinkles lying over $Cu_2O$ nanocrystals are clearly visible in dark contrast. On the other hand, isolated multilayer flakes and thermal, folded wrinkles of different widths can be distinguished in Fig. 4(f), also after oxygen plasma etching. It is also noteworthy that thermal wrinkles are often associated with contamination particles, appearing as dark-contrasted "lentils" with a bright edge.

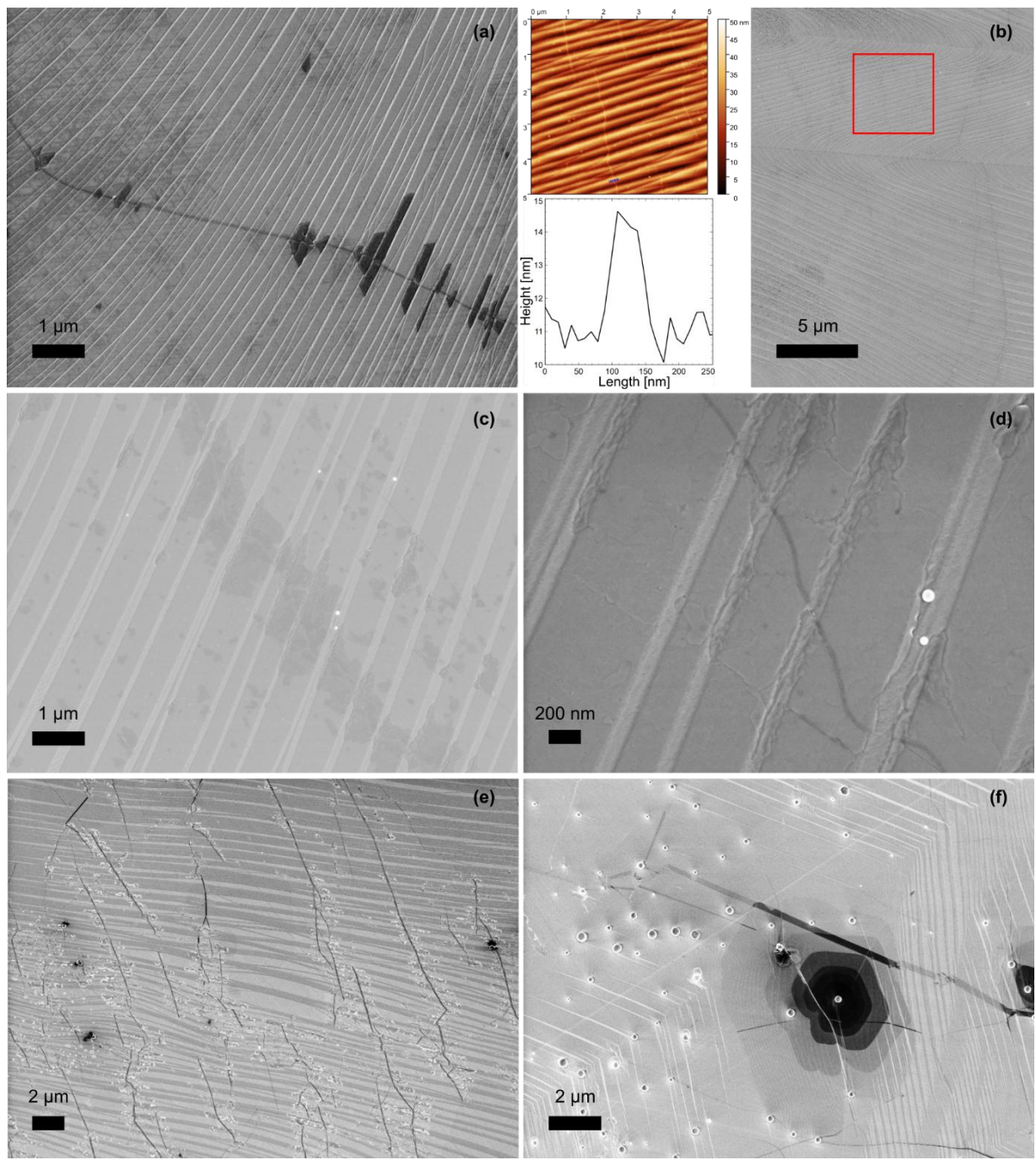

**Figure 4:** (a) High magnification SEM view of a dark-contrasted linear defect in graphene on copper. (b) Right panel: SEM image of graphene on copper after copper wet oxidation showing several dark-contrasted linear defects. Left-panel: AFM topography of the area corresponding to the red square in the right panel and line profile along the blue dotted line, revealing that the dark-contrasted linear defects can be associated to narrow wrinkles. High magnification SEM scans of graphene on copper at the same place (c) before and (d) after oxygen plasma etching. (e) A group of narrow wrinkles on $Cu_2O$ nanocrystals after oxygen plasma etching. (f) Plasma-isolated multilayer flakes and thermal wrinkles of different widths.

The association between copper oxidation and narrow wrinkles means also that water molecules must be able to have access to the underlying copper. This implies that they must be porous, that water can infiltrate through defects to reach copper, and intercalate between copper and graphene.[49] In Fig. 4(a), it can be seen that the oxidation front progresses along copper terraces starting from the copper step edges, which are roughly perpendicular to the narrow wrinkles.[48] Indeed, a nano gap exists between graphene and copper at the step edge, enabling the diffusion of water.[50] From Fig. 3(a), it also appears that the wrinkles that have grown parallel to the copper macro steps correspond to much less oxidized underlying copper, compared to the crossing ones, meaning that the narrow wrinkles do not systematically promote the oxidation of the substrate. Alternatively, a previous work has invoked increased strain to explain enhanced oxidation at the wrinkle location. The authors made the association, although indirectly, between cracks induced in graphene on copper by heating under ambient air on a hot plate and thermal wrinkles.[42] We undertake to replicate that experiment in a direct way by scrutinizing the graphene/copper surface of a fresh sample (they are oxidized coming out of the CVD growth furnace) by SEM before and after thermal treatment at 200 °C for 2 min (longer treatments at 200 °C or at higher temperature simply destroy monolayer graphene). It is found that cracks are indeed revealed or induced in graphene after hot-plate baking (see Figs. S5(a) and (b)) because of the oxidation the copper surface. Still, careful SEM inspection does not allow to link them to anything observable, be it thermal wrinkles or pre-existing cracks (see Figs. S5(c) and (d)), in contradiction with.[42] The cracks seem related to the presence of contamination nanoparticles that may create a structural weak point in graphene and favor the occurrence of cracks after heating in air (see Figs. S5(e)) and the resulting copper oxidation. As reflected in Figs. S5(b) and (e)), they also appear to follow the symmetry axes of the graphene flake (parallel to the edges of the flake, for instance). In Figs. S5(f) and (g), we compare the same region, where thermal wrinkles are clearly identifiable, before and after heating in air, to assess their structural changes. Again, no correlation can be found between cracks and thermal wrinkles.

RS is next conducted on the alleged plasma-isolated wrinkles of Fig. 4 in order to unambiguously confirm the occurrence of graphene at their location and to obtain information on their putative defective nature. Beforehand, the samples have been subjected to a fast SEM scan (to avoid any potential deterioration related to exposure to the electron beam) in order to unequivocally identify wrinkles. The luminescent background of all the spectra presented in Fig. 5 are subtracted. As can be seen in Fig. 5(a), the wrinkles can also be visualized by OM, but only a posteriori. Indeed, since they appear very dim and could be confused with cracks for example, a preliminary SEM scan is necessary to confirm their presence. The image in Fig. 5(a) was captured with the green laser open at low power to prove that the laser spot is indeed located on the narrow wrinkle. A Raman signal pertaining to graphene is clearly visible (D and G bands) in Fig. 5(b), even if, unsurprisingly, with a low intensity. At the same location, the presence of $Cu_2O$ was also found (see Fig. 5(c)), further showing the wrinkle-oxidation coexistence. Contrary to Zhu et al.,[13] we find that the wrinkles are actually defective, in line with.[12,19] This also suggests that wrinkles and other atomic-scale defects (such as pinholes or point defects) very probably play an important role in the oxidation of copper underlying graphene, notably in the case of graphene/copper decoupling. Since the pristine lattice of graphene is watertight,[51] oxidation would take a prohibitively long time if it proceeded only from the edges of the sample towards the center.

One could still argue that the D band could be due to the oxygen plasma treatment and/or to the SEM scan, or to the edges of the plasma-isolated wrinkle.[52] To make sure that the D band originates from the wrinkle itself, we have investigated a pristine (*i.e.* never treated by oxygen plasma) graphene/copper sample (oxidized naturally in air as can be seen in Fig. 5(d)). First, a Raman spectrum of reference monolayer graphene on copper (black dot in Fig. 5(d)), that was preliminarily fast-scanned by SEM, was acquired. The absence of D band in the corresponding spectrum (see Fig. 5(e)) shows that the fast SEM observation has no impact. At the same time, it confirms that copper oxidization underneath graphene does not affect the integrity of graphene. Next, we extract a Raman spectrum of a nearby wrinkle (the 514-nm laser spot is visible in Fig. 5(d))) where a weak D band is detected (see Fig. 5(f)). The corresponding Raman spectrum is much more intense than the one of the plasma-isolated wrinkle in Fig. 5(b). But it could be expected since the Raman signal comes mostly from monolayer graphene, considering the small width of the narrow wrinkle (50-100 nm) compared to the laser spot size of ~1 µm in diameter. Hence, that suggests that the D band in Figs. 5(b) and (f) is intrinsic to wrinkles. It should be acknowledged that this experiment cannot exclude the possibility of defect creation in plasma-isolated wrinkles during the oxygen plasma but it does not at all disprove that "pristine" wrinkles are intrinsically defective. Finally, an AFM scan (together with OM and SEM pictures) of a several plasma-isolated wrinkles is shown in Fig. S6.

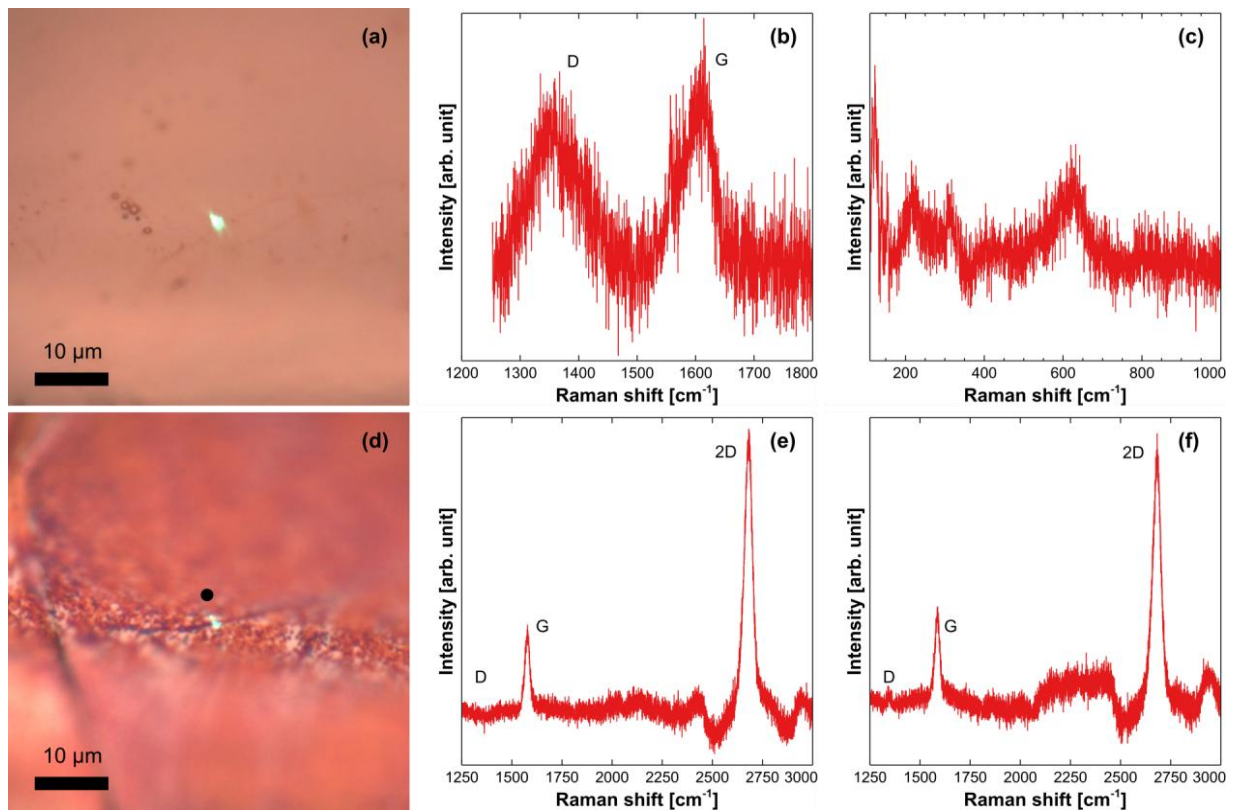

**Figure 5:** (a) OM picture of a plasma-isolated thermal wrinkle on copper. Raman spectra of (b) the corresponding wrinkle (green laser spot in panel (a)) and (c) of associated $Cu_2O$ crystallites. (d) OM picture of a thermal wrinkle on copper. Raman spectra of (e) reference monolayer graphene (black dot in panel (d)) and (f) of the corresponding wrinkle (green laser spot in panel (d)).

In addition to these wrinkles, one can see linear defects with a light-gray contrast (see Fig. 6). If one tries to observe these light-gray features at high SEM magnification to visualize them in more details, the contrast is lost during the scanning procedure because of the deposition of carbon inherent in SEM (see Fig. S7). Nevertheless, it seems clear that these are intrinsic cracks in the as-grown graphene film itself[17,19] since they display a contrast close to the one of copper. Occasionally, we can see cracks parallel to or connected to wrinkles (see Fig. S8), illustrating that they are related to each other. Their morphology (grainy aspect) looks also distinct from the one of narrow wrinkles. These cracks are also different from the reported crack-and-fold defects (since no associated fold is visible).[53] We did not observe such intrinsic cracks in our peculiar case (see Fig. 3(a)), the crack (as well as wrinkle) formation being most likely related to the crystallographic orientation of the underlying copper (see Fig. S9(a)). One can also see oxidation (*i.e.* a dark SEM contrast) along these cracks under graphene, albeit not systematically (see Fig. 6). Also visible with a dark contrast (in the top left corner of Fig. 6) are the well-known, large folded wrinkles. Contrary to the narrow wrinkles studied here above, they do not seem to induce oxidation of copper (as already observed in Fig. 4(f)). Besides, Fig. S9(a-d) illustrates the fact that most of the SEM contrast is lost after the pre-transfer copper wet oxidation in the ethanol/DI water solution, what could be indicative a successful oxidation. However, as already noticed in Fig. 3(b), the color contrast is not lost in the case of OM. Even though the authors who proposed this method applied it to Cu(111) surfaces,[31] it seems also efficient to oxidize other copper crystalline orientations, as already established before in the case of immersion in water.[37] The semi-dry technique was indeed applied successfully several times in our case, on various types of copper orientations given the polycrystalline nature of the copper samples. Another interesting fact to mention is that copper oxidation for samples stored under ambient conditions does not evolve anymore once wet oxidation is performed, regardless of whether copper is covered with graphene or not, contrary to the observations made in Fig. S3 for an untreated sample. This is illustrated in Fig. S10, where Figs. S10(a) and (b) compare the same location before and 6 months after graphene peel-off, respectively, and Figs. S10(c) and (d) compare graphene on copper at the same spot with time interval of 10 months.

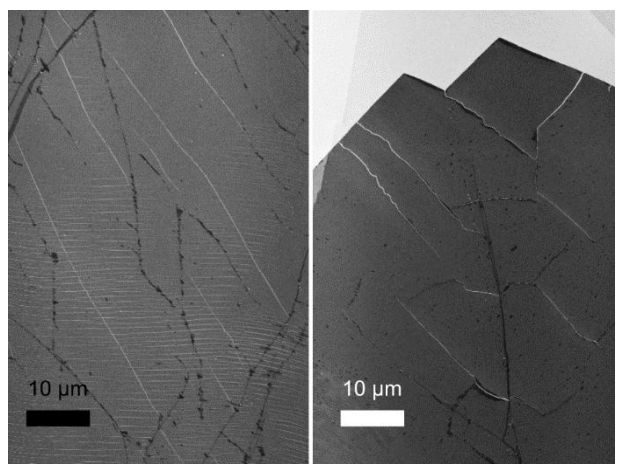

**Figure 6:** SEM pictures of various cracks (bright-contrasted lines) and thermal wrinkles in graphene on copper.

Now that we have established that the dark-contrasted features of Fig. 3(a) are narrow thermal wrinkle, we are going to focus on one clear example (among many others) of each of the three sorts of linear defects detected in Fig. 2(a) to determine their exact nature after transfer onto $SiO_2$/Si. Figure 7(a) highlights, in a high magnification SEM scan (the green rectangle in Fig. 3(a)), an alleged crack in the graphene flake after transfer onto $SiO_2$/Si originating from a narrow wrinkle. A corresponding AFM topographic scan is also provided in Fig. 7(b) which validates this observation unequivocally. The inset to Fig. 7(b) illustrates a line profile of the crack. We can see that the depth of the crack is much greater (~4 nm) than the thickness of a graphene monolayer. This means that graphene must be covered with a thin residual PVA film. The $SiO_2$ film is also covered with a residual layer of PVA both at the place of the crack (see Figs. S11(a) and (b)) and also on the graphene-free zone (see Fig. S11(c)), meaning that PVA cannot be completely stripped by a DI water dip, neither from graphene nor from the $SiO_2$ layer. Note that the AFM scan was performed after taking the SEM image, reason why parallel horizontal lines can be seen in Fig. 7(b) (the imprint of the SEM scan is more visible in an AFM topography scan of a larger surface, see Fig. S11(c) for instance). An additional 24 h long DI water dip or even a 6 h long DI water dip at 80 °C show no improvement whatsoever (see Figs. S12(a-c)), although the same removal method was applied supposedly successfully in previous articles.[26,37,54] The discrepancy might be related to the nature of the PVA (molecular mass, *etc.*) or to the thermal treatment. In contrast, a pristine PVA piece easily dissolves and completely disappears when immersed in DI water, as observed with the naked eye, while it turns out to leave residues after being heated at 105 °C for a few minutes to establish intimate contact with a material, be it graphene or $SiO_2$. Nevertheless, it is surprising that this crack is even observable, as though PVA coating $SiO_2$ is easier to remove than PVA on graphene. However, it seems to be the case since, when we extract a line profile at the edge of the graphene flake, the step is around 3-4 nm as well (see Fig. S11(d)). It can be hypothesized that the chemical affinity of graphene for PVA (and other polymers like PMMA) is greater than for $SiO_2$. In addition, tracing back the crack in graphene to a nanoscale wrinkle is supported by Figs. 7(c) (phase of the AFM scan) and (d) (SEM view) where the linear mark of the same wrinkle and the oxidized copper crystals alongside that wrinkle are clearly apparent. This further demonstrates that narrow wrinkles constitute weaknesses in graphene, as their defective nature was already testified before by the fact that the underlying copper gets oxidized and by RS. The resulting crack is thus plausibly caused by the combination of excessive tensile strain induced by the pressure of the stamp on copper[31] and intrinsic mechanical fragility in graphene at the place of narrow wrinkles, as already evidenced by AFM.[55,56] Nevertheless, in Fig. S13, we can see that the fate of a narrow wrinkle after semi-dry transfer is variable: a narrow wrinkle (the one pinpointed with light-blue dots) does not necessarily crack.

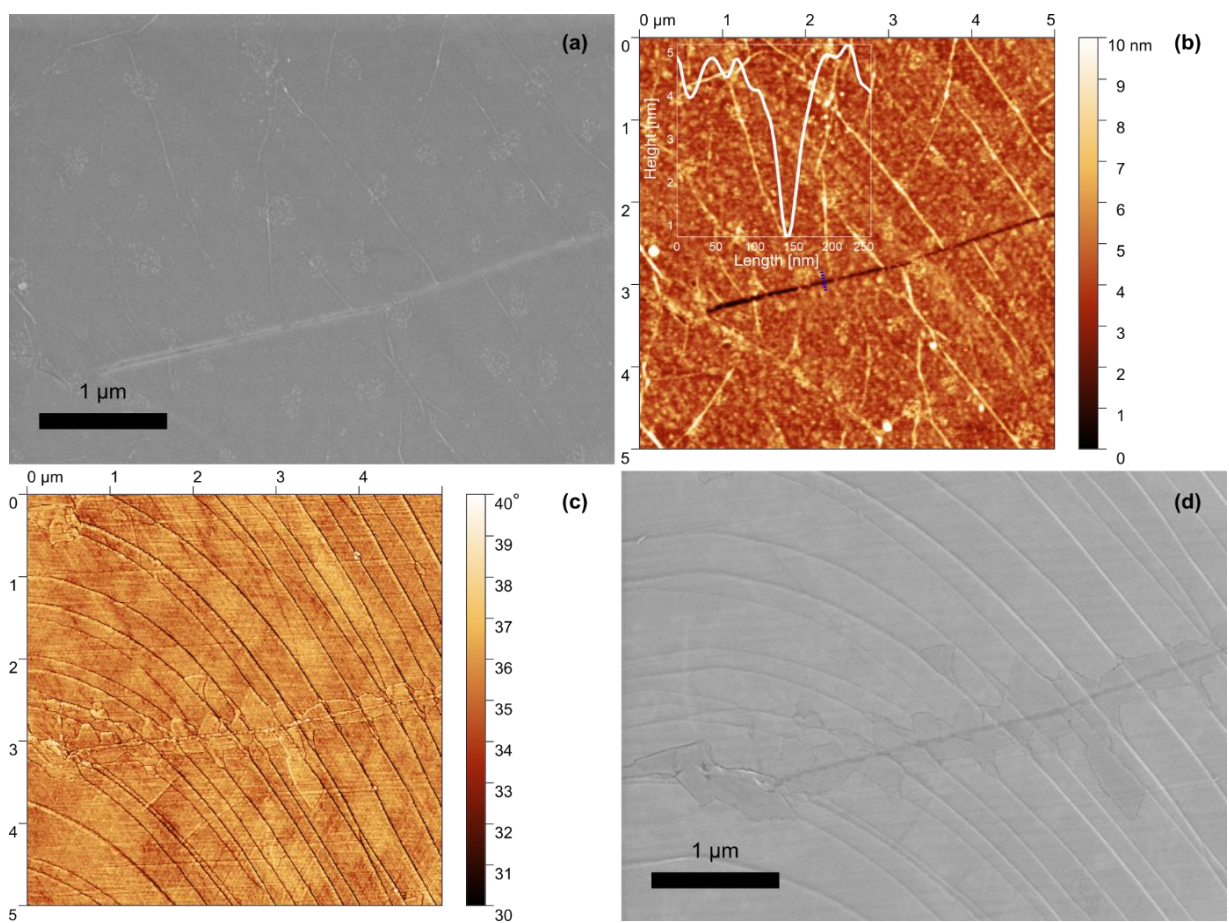

**Figure 7:** (a) High magnification SEM scan and (b) corresponding AFM topography scan of a crack in the graphene flake after semi-dry transfer onto SiO$_2$/Si (green square in Fig. 3(a)). Inset: line profile of the crack corresponding to the blue dotted line. (c) Phase of the AFM scan and (d) corresponding high magnification SEM image of the same location on copper laid bare after graphene pickup.

Figures 8(a) and (b) exhibit an AFM picture of the region evidenced by the red rectangle (second type of linear defects) in Fig. 3(a) after transfer on SiO$_2$/Si and a high magnification SEM picture of the copper foil's surface after graphene pickup of roughly the same area, respectively. From Fig. 8(a), the feature of interest of Fig. 3(a) proves out being a crack. Interestingly enough, a faint imprint of that crack can be observed on the copper surface both in Figs. 2(b) and 8(b) (the yellow frame locating the same region of Figs. 8(a) and (b)). A likely explanation for that trace is that PVA leaves a thin layer of residues on copper after contact (see Fig. S14). This is very convenient for discriminating the origin of the defects and is a unique advantage of the observation of copper after semi-dry transfer. That trace reveals either the presence of a pre-existing crack in graphene (reminiscent of the boundary tears reported in Ref.[17]) or that the crack occurred during the stamping step, even if we could observe no sign of it before transfer in Fig. 3(a). As in the case of the narrow wrinkle, the SiO$_2$ film is covered with PVA at the location of the crack, as can be seen in Figs. 8(c) and (d) exhibiting the AFM topography and phase of the small 1 × 1 µm$^2$ area highlighted by the purple square in Fig. 8(a)).

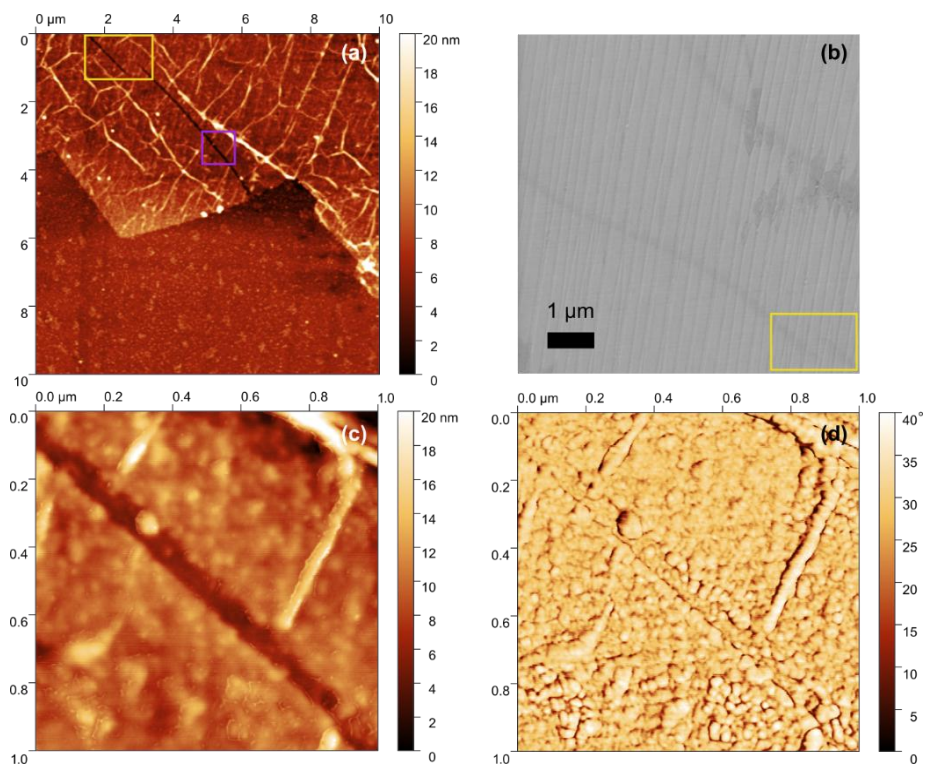

**Figure 8:** (a) AFM topography scan and (b) high magnification SEM view of the region corresponding to the red square in Fig. 3(a) after semi-dry transfer onto SiO$_2$/Si. The yellow rectangle compares exactly the same area in both panels. (c) Small-scale AFM topography and (d) phase scans of the small zone highlighted by the purple square in panel (a).

Next, we make the same kind of analysis for the third case. Figure 9(a) focuses by high magnification SEM imaging on the blue square in Fig. 3(a). The morphology of that feature is indeed strongly suggestive of an extrinsic graphene wrinkle or fold, and this is unequivocally corroborated by the corresponding topographic AFM image in Fig. 9(b). Note again that the trace of Fig. 9(a) can be distinguished in Fig. 9(b) by a rectangle with a weakly darker contrast (and thus lower apparent topography) by the simple fact that the AFM analysis is posterior to the SEM scan, which affects the sample. Contrary to narrow wrinkles, no trace of oxidation can be found on the copper surface after graphene transfer (see Figs. 9(c) and (d) showing the AFM topography and phase of the zone displayed in Fig. 9(b)). Atomic steps crossing each other at a 60° angle (see the yellow equilateral triangle as a visual guide) can be seen in Fig. 9(c). Extrinsic wrinkles also manifest a bright SEM contrast, as opposed to intrinsic ones (see Fig. S15), and show varied heights and widths (insets to Fig. 9). However, the fold is roughly parallel to the copper macro step edges and is very plausibly the outcome of transferring a non-flat graphene film, since it obviously espouses the topography of the copper foil, onto a flat substrate.[19]

To further confirm this, we also place a series of purple dots on the main fold and two other minor folds in Fig. 9(b). They can indeed be matched to macro step edges in Fig. 9(c). We can also see, in Figs. 9(a) and (b), very narrow folds at a 60° angle relative to the main fold (as evidenced by the yellow equilateral triangle in Fig. 9(a)), reflecting the copper crystallographic orientation. But this is not a general rule since many folds also have a random orientation. If one pays careful attention to the AFM topography scans in Figs. 7(b) and 8(a), folds of various heights and widths appear as well as an imprint of the copper macro steps (possibly because residues could be trapped underneath). The widest folds could be the result of the bunching of several narrower folds since we cannot see a fold associated to each copper macro step. One can also observe in Figs. 8(a) and (c) that the crack intersects several folds and that these folds are not continuous on either side of the crack. This is a clue that the folds are induced by the transfer process and were not present before. Note finally that heating at 80 °C for 6 h does not help at all in suppressing or attenuating the folds, as illustrated in Fig. S12. All of this plainly illustrates that the folds arise from the semi-dry transfer, in relation with the topography of copper.

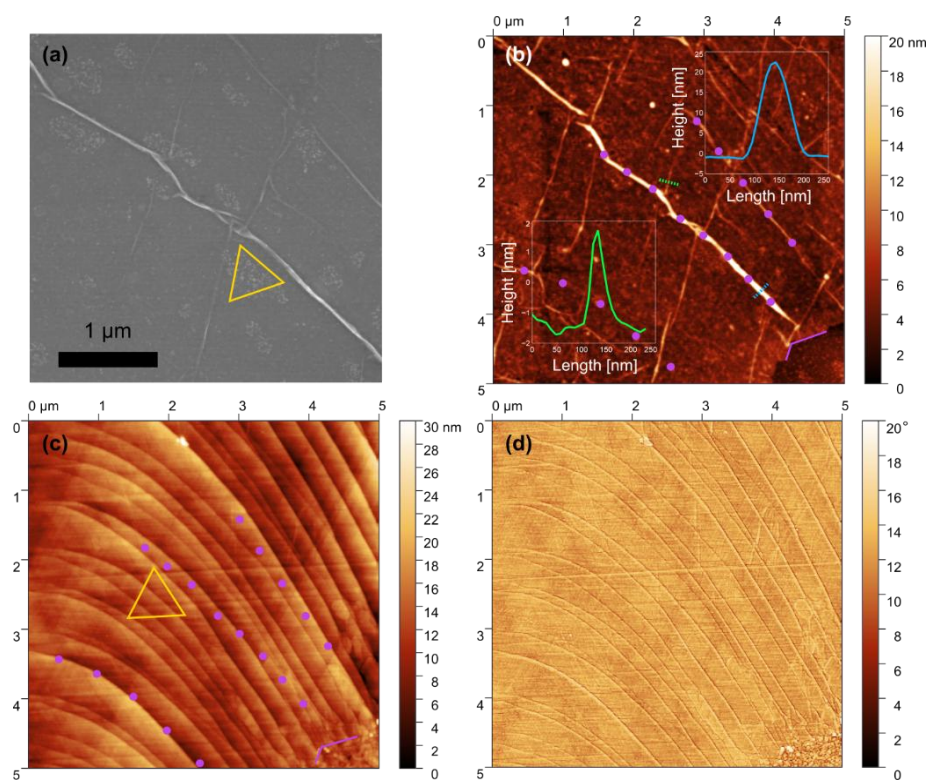

**Figure 9:** (a) High magnification SEM micrograph and (b) AFM topography scan of a graphene fold after semi-dry transfer onto SiO$_2$/Si (blue square in Fig. 3(a)). The insets are line profiles of the main fold (blue dotted line and blue curve) and of a minor one (green dotted line and green curve). (c) AFM topography and (d) phase scans of the same zone on copper, exposed after graphene pickup.

Actually, we can even pinpoint a fourth situation, hybrid between the first and second case that could in fact already be partially evidenced in the top right corner of Fig. 8(b). We can see the superposition of Cu$_2$O nanocrystals left over by narrow thermal wrinkles (as in the first case) and a PVA trail (as in the second case), also resulting in cracks after semi-dry transfer (see Fig. S16 for a more complete view). This overlap of a PVA trail on Cu$_2$O nanocrystals further suggests that the crack is indeed created on the copper foil, before picking up graphene, as already hypothesized in the first case. An additional illustration of the first and fourth cases are given in Fig. S17. Moreover, as one can notice in Fig. S18, more often than not, the PVA trace does not necessarily exactly overlap the crack as if the stamp is sliding slightly during the transfer while pressing the stamp on the sample. It also shows that this hybrid situation is quite common and that it actually encompasses the first case.

All in all, the previous observations illustrate that the situation is rather complex. Figure 10 summarizes the three cases that were identified before. We have focused our attention on a region in particular, with a given copper orientation. Things would certainly be different on other copper crystalline orientations but the same kind of analysis can apply. The main lesson to learn from this investigation is that wrinkles are defective and can result in cracks after transfer, supplementary reason why they must be suppressed, besides the fact that they create thickness inhomogeneities in graphene. Furthermore, cross-correlation between different characterization techniques must be performed to avoid drawing incorrect conclusions considering that the nanoscale size of the observed features. This is exemplified by SEM, which can affect the surface under examination (and notably result in contrast loss) and whose contrast can be misleading and depends on the nature of the substrate (crystalline orientation, conductivity, *etc.*).

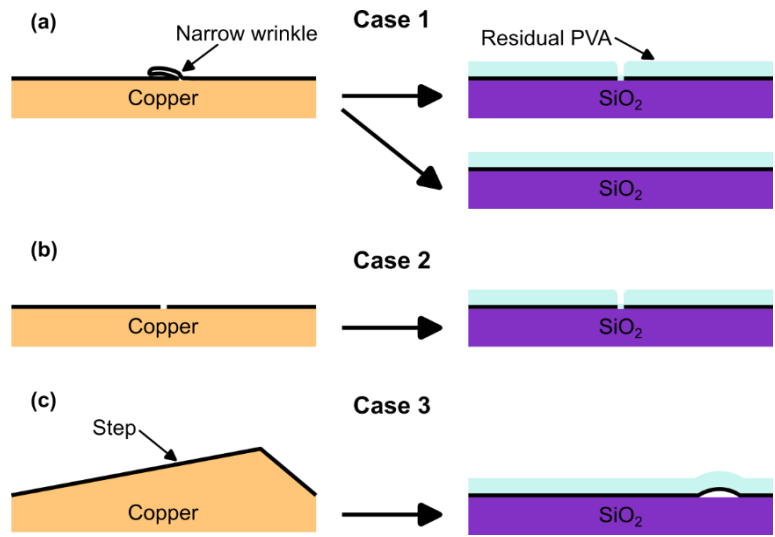

**Figure 10:** Schematic summary of the three identified categories of linear defects.

**3.2) Microscale defects**

In this part, we investigate the quality of the transferred graphene flake at a larger scale by resorting to OM as a fast inspection technique. Figures 11(a) and (b) confront a larger portion of the graphene flake observed on $SiO_2$/Si and on copper foil, after graphene pickup, respectively. The contour of the flake is evidenced with a black dotted line in Fig. 11(b). The graphene sheet is mostly monolayer, with very few multilayer inclusions (showing that multilayers can be transferred by the technique), while several holes can be seen here and there in Fig. 11(a). The two regions highlighted by a red rectangle in Fig. 11(a) correspond to two graphene areas that remained stuck to the copper foil after transfer (the zone corresponding to Fig. 3 is comprised in the green frame. Upon close inspection, they can indeed be seen in Fig. 11(b), even if faintly, since they display a slightly darker orange contrast compared to the bare copper surface. For clarification, higher magnification views of the same areas are shown in Figs. S19(a) and (b). In Fig. S19(c), by using a simple thresholding procedure (with the ImageJ software[57]) on Fig. 11(b) to extract the contour of the main topographic features of the copper foil (grain boundaries and rolling striations), we superimpose the obtained image (with the background made transparent) on Fig. 11(a). To make a correct alignment of both images, we have chosen a few easily recognizable reference points. As could be expected, the holes in graphene can be fairly well correlated with the copper foil topography. Notably, that the extracted contour does not perfectly overlap Fig. 11(a) even though the hole-topography correlation is clear. This can be explained by the fact that, since the copper foil is not flat, neither is the graphene film grown on top of it. When transferred to the flat surface of the $SiO_2$/Si substrate, graphene is stretched and can break in places corresponding to bumps (rolling striations) and troughs (copper grain boundaries). In addition, since the procedure is entirely manual (with a poor control of the pressure applied to the stamp), it is certainly tougher than the traditional polymer-assisted wet transfer. In Figs. 11(c) and (d), we focus on a smaller region of Figs. 11(a) and (b). The correspondence between the topography and punctured graphene areas becomes even more blatant (grains boundaries and one rolling striation obtained from Fig. 11(c) featured by black dotted curves are overlapped on Fig. 11(d)). In the inset to Fig. 11(d), exhibiting an SEM image (with some degree of transparency) of the area on copper after transfer, residual graphene fragments coinciding with copper grain boundaries are evidenced. Additionally, Fig. 11(e) discloses a representative Raman spectrum obtained for graphene on $SiO_2$/Si. This spectrum is typical of monolayer graphene, with a G band at 1584 cm$^{-1}$, a 2D band at 2688 cm$^{-1}$ and full width at half maximum of 29.5 cm$^{-1}$, and a weak D band at 1350 cm$^{-1}$. The D to G peak intensity ratio, $I_D/I_G$, amounts to 0.087, corresponding to a distance between zero-dimensional pointlike defects, $L_D$, of 38 nm.[58] This firmly places this CVD graphene in the low defect regime ($L_D$ > 10 nm) and testifies to its high quality. Additional quantitative data are also available in Fig. S20(a)). The Raman peak that can be spotted at 2330 cm$^{-1}$ corresponds to atmospheric molecular $N_2$.[59,60] It may become visible after acquisition at long integration times.[61] The spectrum is acquired in a region far from the graphene area inspected here above, because of the degradation caused by its consequent SEM inspection, as testified by the occurrence of a significant D band at 1350 cm$^{-1}$ altogether with a luminescent background (see Fig. S20(b)). In addition, no peak relevant to PVA is detected in spite of the presence of a residual PVA overlayer, the most intense PVA Raman peak centered being centered at 2900 cm$^{-1}$ and spanning between 2800 and 3000 cm$^{-1}$ (see Fig. S20(c)).

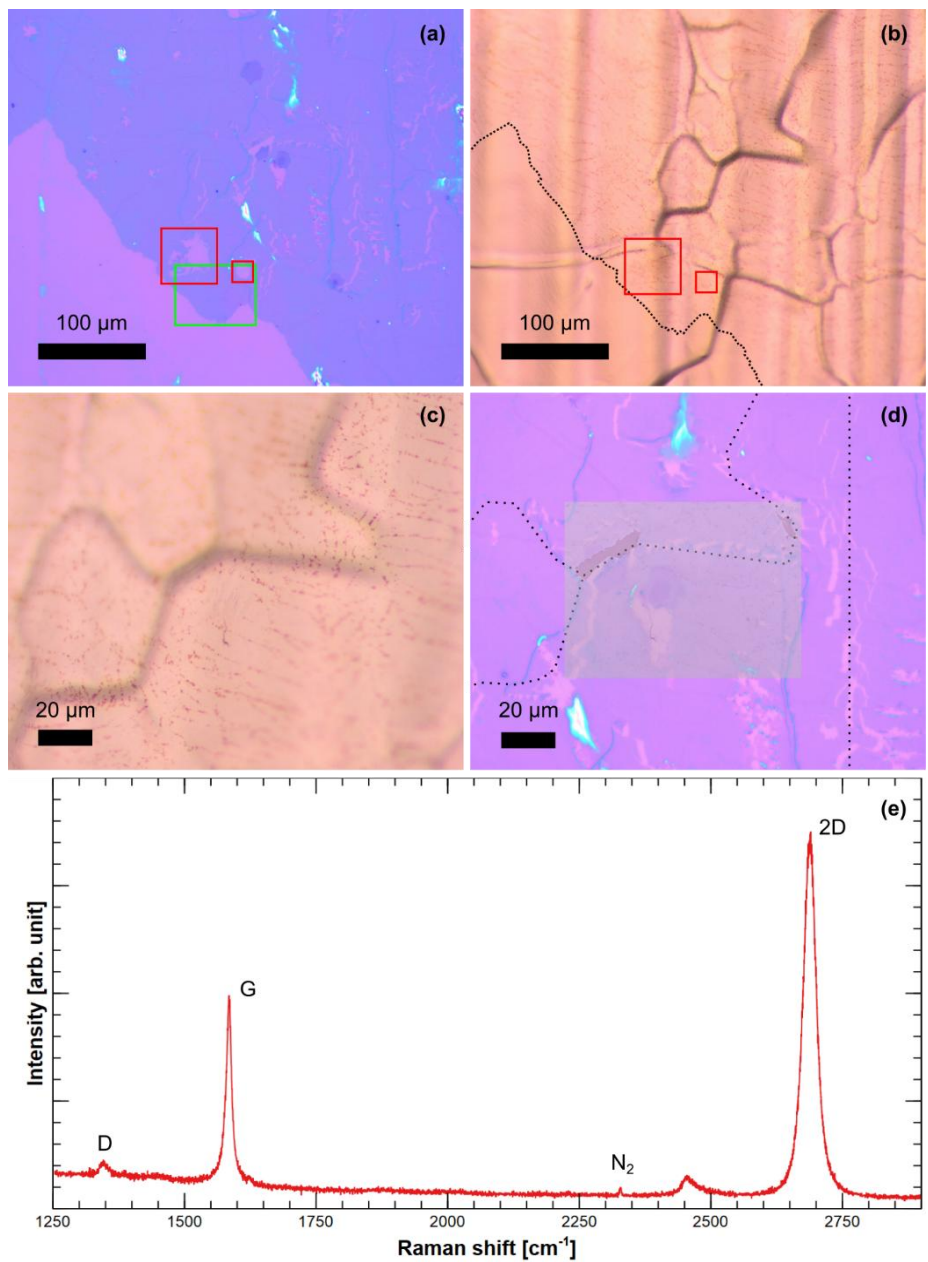

**Figure 11:** (a) Larger part of the semi-dry transferred graphene flake observed by OM on SiO$_2$/Si and (b) the corresponding zone on copper foil after graphene pickup. The two red frames indicate graphene that was not lifted off the copper foil. The green rectangle emphasizes the zone imaged in Fig. 3. Zoom-in on a smaller portion (c) of panel (b) and (d) of panel (a). A SEM picture taken on copper after graphene transfer at the same scale is superimposed on panel (d). (e) Raman spectrum of monolayer graphene after semi-dry transfer onto SiO$_2$/Si.

Finally, it is worthwhile scrutinizing the same region at the highest magnification (100×) attainable with our optical microscope, in order to assess if we can deduce anything about the nanoscale defects tackled in the previous section with OM alone, thus dispensing with a slow technique like AFM or causing damage such as SEM. To do so, we compare an OM picture (see Fig. S21(a)) to corresponding SEM scans on SiO$_2$/Si (see Fig. S21(b)), and of the copper surface after and before transfer (see Figs. S21(c) and (d)), respectively. In Fig. S21(a), we can clearly distinguish narrow wrinkles and cracks (although these are more difficult to spot). As opposed to OM, wrinkles and multilayers are poorly contrasted in SEM (see Fig. S21(b)), probably because of the residual PVA thin film, while the cracks are clearly visible. They can be related to their counterpart on copper (see Figs. S21(c) and (d)), confirming again the fact that narrow wrinkles on copper can break during the transfer, but not systematically. Some wrinkles and cracks are selected and highlighted in Fig. S21(a), and corresponding trails are superimposed on Figs. S21(b-d) (and slightly rotated if necessary). The corresponding tracks can be very well mapped to the dark-contrasted features on copper when they belong to a single copper grain while they can be weakly shifted when separated by a grain boundary where graphene broke since the two graphene pieces can slide relatively to each other on either side of the boundary, as already noticed before in Fig. S19(c).

## 4) Conclusion

In this study, we used correlated OM, SEM, and AFM analyses of graphene — before and after semi-dry transfer onto $SiO_2$/Si, as well as the underlying copper post-transfer — to identify the origin of various defects. Most defects, except for micrometer-scale holes and tears, are nanoscale and one-dimensional (cracks, wrinkles, folds). Our observations directly link copper surface morphology to specific graphene defects at different transfer stages, clarifying their origin. Although focused on a single copper orientation, our findings suggest a broader applicability. Future studies using electron backscatter diffraction could further explore the influence of copper crystallography, particularly on crack formation, which remains underreported. This work lays the groundwork for such investigations.

## Supporting Information

Supporting Information is available from the Wiley Online Library or from the author.

## Acknowledgements

B. H. (Senior Research Associate) acknowledges financial support from the F.R.S.-FNRS (Belgium). The present research was funded by the European Union's Horizon Europe research and innovation program under grant agreement No. 101099139 ("FLATS" project).B. H. (Senior Research Associate) acknowledges financial support from the F.R.S.-FNRS (Belgium). The present research was funded by the European Union's Horizon Europe research and innovation program under grant agreement No. 101099139 ("FLATS" project).

## Conflict of Interest

The authors declare no conflict of interest.

## Data Availability Statement

The data that support the findings of this study are available in the Supporting Information of this article.

## Keywords

Chemical vapor deposition, graphene, wrinkles, cracks, semi-dry transfer.